\documentclass[11pt, letterpaper]{amsart}

\usepackage[parfill]{parskip}
\usepackage{amssymb}
\usepackage{amsmath}
\usepackage{amsthm}
\usepackage{amsfonts}

\usepackage[T1]{fontenc}
\usepackage{palatino}
\usepackage[text={6.5in, 9in}, centering]{geometry}

\usepackage[usenames]{color}
\newtheorem{theorem}{Theorem}
\newtheorem{definition}[theorem]{Definition}

\newtheorem{remark}[theorem]{Remark}

\newcommand{\R}{\mathbb{R}}

\newcommand{\E}{\mathcal{E}}
\newcommand{\supp}{\mbox{supp}}

\title[Concentration of enstrophy in 3D MHD]{On the transport and concentration of enstrophy in 3D magnetohydrodynamic turbulence.}
\author{Z. Bradshaw and Z. Gruji\'c}
\address{Department of Mathematics\\
University of Virginia\\ Charlottesville, VA 22904}
\date{\today}

\begin{document}
\begin{abstract}Working directly from the 3D magnetohydrodynamical equations and entirely in physical scales we formulate a scenario  wherein the enstrophy flux exhibits cascade-like properties.  In particular we show the inertially-driven transport of current and vorticity enstrophy is from larger to smaller scale structures and this inter-scale transfer is local and occurs at a nearly constant rate. This process is reminiscent of the direct cascades exhibited by certain ideal invariants in turbulent plasmas.  Our results are consistent with the physically and numerically supported picture that current and vorticity concentrate on small-scale, coherent structures.
\end{abstract}
\maketitle

\section{Introduction.}

The magnetohydrodynamic equations (MHD) model the evolution of a coupled system comprised of a magnetic field and a velocity field associated with an electrically conducting fluid.  
Throughout turbulent MHD regimes, observational and numerical data support a picture of intermittently distributed regions of high spatial complexity -- the coherent current and vortex structures; i.e. current sheets -- which become increasingly thin as turbulence evolves \cite{DB, Greco09, Greco08, Bruno2001,PoBhGa}. One view is that this process is initiated and driven by inertial effects associated with the cascades of ideally conserved quantities (see \cite{MaMoWaSe-Review,Eyink2006,Servidio_et_al_2011,Zhdankin_et_al_2012}; regarding the genesis of solar wind current sheets, an alternative view is given in \cite{Bruno2001,Borovsky08} where it is speculated that they are magnetic flux tubes generated in the solar corona and passively advected by the solar wind; modern theories and numerical studies regarding turbulent cascades and spectra can be found in \cite{B05,BhaPo10,DB-ES-01,PeBo10} and are discussed briefly below).  Although enstrophy -- taken to mean the sum of the squares of vorticity and current -- is not conserved in the ideal equations (and so we avoid the term ``enstrophy cascade''), it does become increasingly concentrated on small-scale current sheets.   Understanding the concentrative process is important because it effects the development of current-driven instabilities (for example, the tearing instability; cf. \cite{DB}) which drive magnetic reconnection (the relationship between magnetic reconnection and current structures has been studied in \cite{Eyink2006,PoBhGa}).   This paper is concerned with establishing conditions which rigorously affirm that the inertially driven transport of enstrophy is concentrative.  In particular, we show the enstrophy flux is predominantly oriented (in a statistical sense) from larger to smaller scale structures and, moreover, is local and occurs at a nearly constant rate.  In the case that non-inertial effects are negligible at large scales this indicates a detectable concentration of enstrophy where the inertial contribution is cascade-like.  Also of note is the implication that, at least in our scenario, inertial forces -- i.e. advection of the enstrophy by the fluid --  effect the morphology of current sheets and, therefore, these structures are not passively advected by the fluid medium.

Although our main interest is the 3D setting, for illustrative purposes it is useful to consider the concentrative progression in 2D (we here summarize \cite{DB}; it is the intention of the authors to comment specifically on the 2D case in a future paper concerned with a variety of features of 2D MHD turbulence including the inverse cascade of magnetic potential and the direct cascade of energy).  In 2D, early in the turbulent evolution, the current is predominantly distributed along eddy boundaries while, at later stages, it becomes spiked at eddy centers.
Unlike 2D fluid turbulence where the concentration of enstrophy is driven solely by inertial forces -- the enstrophy cascade -- the Lorentz force does not vanish in 2D MHD and therefore introduces a possible source of enstrophy \cite {Chorin,Frisch05,Co94,CoPrSe-95}.  This complicates the creation, transformation, transport, and destruction of enstrophy and, consequently, something superficially similar to the fluid ``enstrophy cascade'' should only be visible if the Lorentz force is depleted by some intrinsic mechanism and the enstrophy flux is nearly constant, local, and directed from larger to smaller scales.  Heuristic arguments and numerical evidence indicates that such depletive effects exist (cf. \cite{kinney95,DB,DB-ES-01}) and this supports the idea that, at least in 2D, enstrophy undergoes a concentrative process with qualities similar to a cascade. In 3D MHD the situation is further complicated by stretching effects which can contribute to or detract from the concentrative process.  Indeed, it is possible for the concentration to occur independently of inertial effects if enstrophy is depleted at large scales and sourced at small scales (this is why the concentrative process is distinct from the notion of ``cascade''). Our interest is whether or not, in a circumstance where non-inertial effects experience depletion, inertial effects on enstrophy are concentrative.

Our work is closely related to the inertial effects which drive the energy cascade and it is appropriate to remark on the status of this field. The existence of an energy cascade in 3D MHD is widely accepted (see \cite{DB} for an overview and \cite{Kr,Ir} for the classical phenomenologies) and has recently been rigorous established in \cite{BrGr1} as an intrinsic feature -- i.e. independent of assumptions regarding the geometry or strength of the mean magnetic field -- of decaying MHD turbulence.  Although existence is granted, the specifics of the turbulent spectrum remain contentious.
Since the work of Goldriech and Sridhar (cf. \cite{GS94,GS95}), attention has focused on the anisotropy exhibited by turbulent plasmas in the presence of a strong magnetic mean field (this contrasts the classical phenomenology of Iroshnikov and Kraichnan which assumed an isotropic spectral transfer; cf. \cite{Kr,Ir}).  In \cite{GS95}, based on the \emph{critical balance} assumption -- i.e. that there is a single timescale for parallel and perpendicular motion (to the magnetic mean field) in a turbulent eddy -- a distinct perpendicular energy spectrum was derived and a scaling relationship established between the lengths of perpendicular and parallel fluctuations.  Subsequent numerical results indicated a more complex picture than that in \cite{GS95} and various competing phenomenologies have since been developed to account for this (cf. \cite{B05,MuGr05,GaPoMa05,BeLa08}). Lively debate remains as to which is the most effective \cite{PeMaBoCa12,Be12}.  One of these (cf. the phenomenology of Bolydrev;  \cite{B05}) considers the depletive effect of a proposed \emph{dynamic alignment} -- the tendency of magnetic and velocity orientations to become progressively aligned at smaller scales -- on non-linear interactions and results in an additional anisotropy in the plane perpendicular to the mean magnetic field which is consistent with the formation of small-scale current sheets (as opposed to filamentary structures).

The present work is primarily interested in two conclusions: (1) that the enstrophy flux is predominantly oriented from large to small scale structures across a range of scales and (2) that the enstrophy flux is nearly constant across this  range. These indicate that the inter-scale transport of enstrophy is cascade-like (even if the apparent evolution of enstrophy does not display this),  which is consistent with the concentrative picture highlighted above provided additional non-linear effects -- creation, stretching, or dissipation of enstrophy -- are depleted by current sheet geometry or positively contribute only at small-scales.  This also indicates that the inertial transport of enstrophy is \emph{active} in that fluid advection effects the structure of current sheets.

To achieve these conclusions we use a dynamic, multi-scale averaging process developed to study features of hydrodynamic turbulence (we recall the specifics of this methodology in Section 2; see also \cite{DaGr1,DaGr3}).  The process acts as a detector of significant \emph{sign-fluctuations} associated with a physical density at a given scale and is used to show that the orientation of a particular flux -- i.e. the enstrophy flux -- is, in a statistically significant sense, from larger to smaller scales, thereby indicating a concentration effect toward structures of progressively fine scale.
To achieve this we will establish several dynamic estimates for quantities associated with non-inertial terms (originating in the current-vorticity formulation of 3D MHD; see Sections 4 and 5) and we include several assumptions (see (A1)-(A3) in Section 3) to accommodate these estimates. Chief among these is a requirement, (A1), that the vorticity field satisfies a hybrid geometric/smoothness property in the region of high spatial complexity.  Requirements of this type (i.e. conditions depleting non-linearities) have been used to formulate conditional regularity results for both 3D NSE (cf. \cite{CoFe93,BeBe}) and 3D MHD (cf. \cite{Wu2002,Wu00,HeXin2}) and our particular formulation is chosen for its robustness with regard to its mathematical applications. We include a comment highlighting a more complicated but potentially more physically motivated configuration with the same mathematical effects (see Remark \ref{rem:assumptions} following our statement of assumption (A1)). 

Scale-locality of the enstrophy flux is included as a direct corollary of our main result (see Section 6).  Even for cascading quantities, the locality question is generally more complicated in 3D MHD than in 3D NSE due to the variety of transporting mechanisms (cf. \cite{AlDiss,AlEy10}).  Under the same conditions that indicate the inertial transfer of enstrophy is concentrative, we affirm that the inertially driven inter-scale transfer is predominantly between comparable scales and, moreover, this locality propagates exponentially along the dyadic scale.

\section{$(K_1,K_2)-$Covers and Ensemble Averages.}

The main purpose of this section is to describe how {\em ensemble averaging} with respect to {\em $(K_1,K_2)$- covers} of an integral domain $B(0,R_0)$ can be used to establish the \emph{essential positivity} of a potentially sign-varying density over a range of physical scales associated with the integral domain.  The application to turbulence is establishing the positivity of certain inward directed flux densities -- i.e. that inter-scale transfer is uni-directional from larger to smaller scales indicating the transfer is concentrative -- as well as the near-constancy of the averaged densities -- i.e. the space-time averages over cover elements are all mutually comparable -- across a range of scales.

The ensemble averages will be taken over collections of spatio-temporal averages of physical densities localized to cover elements of a particular type of covering -- a so called {\em $(K_1,K_2)$-cover} -- where the cover is over a macro-scale region where turbulent activity is evident.  For simplicity, this region will be taken as a ball of radius $R_0$ centered at the origin and is referred to as the {\em integral domain}.
The $(K_1,K_2)$-covers are now defined.

\begin{definition}Let $K_1,K_2\in \mathbb N$ and $0\leq R\leq R_0$.  The cover of the integral domain $B(0,R_0)$ by the $n$ balls, $\{B(x_i,R)\}_{i=1}^n$ is a  {\em $(K_1,K_2)$-cover at scale }$R$ if,
\begin{align*}\bigg( \frac {R_0} R \bigg)^3\leq n \leq K_1 \bigg(\frac {R_0} R\bigg)^3,\end{align*}
and, for any $x\in B(0,R_0)$, $x$ is contained in at most $K_2$ balls from the cover.
\end{definition}

In the hereafter all covers are understood to be $(K_1,K_2)$- covers at scale $R$.
The positive integers $K_1$ and $K_2$ represent the maximum allowed \emph{global} and \emph{local multiplicities}, respectively.

In order to localize a physical density to a cover element we incorporate certain {\em refined} cut-off functions.  For the cover element centered at the point $x_i\in \R^3$, let $\phi_i(x,t)=\eta(t)\psi(x)$ where $\eta\in C^\infty(0,T)$ and $\psi\in C_0^\infty (\R^3)$ satisfy,
\begin{align}0\leq \eta\leq 1,\qquad \eta=0~\mbox{on }(0,T/3),\qquad\eta=1~\mbox{ on }(2T/3,T),\qquad\frac {|\partial_t\eta|} {\eta^\delta }\leq \frac {C_0} T,
\end{align}
and,
\begin{align}\label{spacecutoff} 0\leq \psi\leq 1,\qquad\psi=1\mbox{ on }B(x_i,R),\qquad\frac {|\partial_i\psi|} {\psi^{\rho}} \leq \frac {C_0} {R},\qquad\frac {|\partial_i\partial_j \psi|} {\psi^{2\rho-1}}\leq \frac {C_0} {R^2},
\end{align}where $3/4< \delta,\rho <1$ and $C_0$ is a fixed constant.

By $\phi_0$ we denote the spatially radial cut-off function associated with the integral domain --  the ball centered at $0$ of radius $R_0$ -- satisfying the above properties and supported in $B(0,2R_0)$.

Comparisons will be necessary between averaged quantities localized to cover elements at some scale $R$ and averaged quantities at the integral scale.  To accommodate this we impose several additional conditions on $\psi$ when the center of the associated cover element is near the boundary of $B(0,R_0)$.  If $B(x_i,R)\subset B(0,R_0)$ we assume $\psi\leq \psi_0$ and $\supp ~\psi \subset B(x_i,2R)$.  Alternatively, when $B(x_i,R)\not\subset B(0,R_0)$ we stipulate that $\psi=1$ on $B(x_i,R)\cap B(0,R_0)$, satisfies (\ref{spacecutoff}), and we additionally have,
\begin{quote}$\psi=\psi_0$ on the intersection of $S(0,R_0,2R_0)$ and the cone with apex at the origin and with boundaries passing through the intersection of the circle centered at the origin of radius $R_0$ and the boundary of $B(x_i,R)$,
\end{quote}
and,
\begin{quote}$\psi=0$ on the intersection of the three sets $B(0,R_0)\setminus B(x_i,2R)$, $S(0,R_0,2R_0)$, and the outside of the cone with apex at the origin and boundaries passing through the intersection of the circle centered at the origin of radius $R_0$ and the boundary of $B(x_i,2R)$.
\end{quote}
With these stipulation it is clear that $\phi_i$ can be constructed so that $\phi_i\leq \phi_0$ and the gradients of $\phi_i$ are inwardly oriented.

The above apparatus is employed to study properties of a physical density at a {\em physical scale} $R$ associated with the integral domain $B(0,R_0)$ in a manner which we now illustrate.  Let $\theta$ be a physical density (e.g. a flux density) and define its localized spatio-temporal average on a cover element at scale $R$ around $x_i$ as, \begin{align*}\tilde \theta_{x_i,R} = \frac 1 T \int_0^T \frac 1 {R^3} \int_{B(x_i,2R)} \theta(x,t)\phi^\delta_{i}(x,t)~dx~dt,\end{align*}
where $0< \delta\leq 1$, and let $\langle \Theta\rangle_R$ denote the ensemble average over localized averages associated with cover elements,
\[\langle \Theta \rangle_R=\frac 1 n \sum_{i=1}^n \tilde \theta_{x_i,R}.\]
Examining the values obtained by ensemble averaging the averages associated to a variety of covers at a fixed  scale allows us to draw conclusions about the flux density $\theta$ at comparable and greater scales.  For instance, stability away from zero (i.e. near constancy) of $\{\langle \Theta\rangle _R\}$ indicates that the sign of $\theta$ is essentially uniform at scales comparable to or greater than $R$.  On the other hand, if the sign were not essentially uniform at scale $R$, particular covers could be arranged to enhance negative and positive regions and thus give a wide range of sign varying values in $\{\langle \Theta\rangle _R\}$.  In order, then, to show the essential positivity of an {\em a priori} sign varying density $\theta$ at a scale $R$ it is sufficient to show the positivity and stability of $\{\langle \Theta\rangle _R\}$.

Conversely, if $\theta$ is an {\em a priori} non-negative density, then ensemble averages are all comparable to the integral scale average across the range $0< R\leq R_0$.  Precisely put, there exists $K_*$ depending only on $K_1$ and $K_2$ so that,
\begin{align}\label{PosDensityInterp}\frac 1 {K_*} \Theta_0\leq \langle \Theta\rangle _R\leq K_*\Theta_0,
\end{align}where,
\begin{align*}\Theta_0=\frac 1 T \int_0^T \frac 1 {R^3} \int_{B(0,2R_0)} \theta(x,t)\phi_0^\delta(x,t)~dx~dt.\end{align*}
The inequalities \eqref{PosDensityInterp} follow in particular from the selection of our cut-off functions as well as the defining properties of our cover.  Indeed, by directly comparing the integrands of the involved localized spatio-temporal averages to the integral domain spatio-temporal average, one sees that,\begin{eqnarray}\label{ineq:interp}\frac 1 {K_1}\Theta_0 \leq \langle \Theta\rangle _R\leq K_2 \Theta_0.
\end{eqnarray}
For additional discussion of $(K_1,K_2)$-covers and ensemble averages, including some computational illustrations of the process, see \cite{DaGr3}.

\section{Enstrophy Concentration.}

Our mathematical setting is that of weak solutions to the magnetohydrodynamic equations (cf. \cite{SeTe} for the essential theory).  Define $\mathcal V=\{f\in C_0^\infty (\R^3):\nabla\cdot f=0~\mbox{in the sense of distributions}\}$ and let $H$ be the closure of $\mathcal V$ under the norm of the $(L^2(\R^3))^3$. By a solution to 3D MHD we mean a weak (distributional) solution to the following coupled system (3D MHD),
\begin{align*}u_t-  \triangle u +(u\cdot \nabla)u- (b\cdot \nabla)b +\nabla P &=0,
\\ b_t- \triangle b + (u\cdot \nabla) b - (b\cdot \nabla )u &=0 ,
\\ \nabla\cdot u = \nabla\cdot b&= 0,
\\ u(x,0)=u_0(x)&\in H,
\\ b(x,0)=b_0(x) & \in H,
\end{align*}
where the magnetic resistivity and kinematic viscosity have been set to one and $P$ is the total pressure.  We additionally require that $u$ and $b$ are locally smooth.

Taking the curl of the above equations yields the following evolution equations for the vorticity and current, denoted $\omega$ and $j$ respectively,
\begin{align}\label{eq:vorticity}\partial_t \omega-\Delta \omega&=-(u\cdot \nabla)\omega+(\omega\cdot \nabla)u+(b\cdot \nabla)j-(j\cdot \nabla)b,
\\ \label{eq:current}\partial_t j-  \Delta j &=-(u\cdot \nabla)j+(j\cdot \nabla)u+(b\cdot \nabla)\omega-(\omega\cdot \nabla)b+2\sum_{l=1}^3 \nabla b_l \times \nabla u_l.
\end{align}
In our study we substitute for the inward kinetic and magnetic enstrophy fluxes through the boundary of a ball, $B=B(x_0,2R)$,
\begin{align*}-\int_{\partial B}\frac 1 2 |\omega|^2 (u\cdot n)~d\sigma &=-\int_B(u\cdot \nabla)\omega \cdot \omega ~dx,
\\ -\int_{\partial B} \frac 1 2 |j|^2 (u\cdot n)~d\sigma &= -\int_B(u\cdot \nabla)j \cdot j ~dx,\end{align*}
the inward kinetic and magnetic enstrophy flux through a shell, $S(x_0,R,2R)$, by incorporating a (nearly radial) cut-off function, $\phi$.  This cut-off function was defined in Section 2 and is chosen so that the gradient is directed inward.  After multiplying $(u\cdot \nabla)~\omega$ and $(u\cdot \nabla)~j$ respectively by $\phi ~\omega$ and $\phi~ j$, we have the following realization of the {\em local kinetic and magnetic enstrophy fluxes at scale $R$ around the point $x_0$},
\begin{align*}\Phi^\omega_{x_0,R}&:=\int \frac 1 2 |\omega|^2(u\cdot \nabla \phi )~dx=-\int (u\cdot \nabla)\omega\cdot(\phi \omega)~dx,
\\ \Phi^j_{x_0,R}&:=\int \frac 1 2 |j|^2(u\cdot \nabla \phi )~dx=-\int (u\cdot \nabla)j\cdot(\phi j)~dx,
\end{align*}
and we define the local \emph{combined enstrophy flux} by $\Phi_{x_0,R}=\Phi^\omega_{x_0,R}+\Phi^j_{x_0,R}$.
Formulas for the localized enstrophy fluxes are realized via the non-linear terms $(u\cdot \nabla)~\omega$ and $(u\cdot \nabla)~j $ by multiplying \eqref{eq:vorticity} and \eqref{eq:current} respectively by $\phi~ \omega$ and $\phi ~j$ and integrating. In this manner we see that the localized kinetic enstrophy flux is given by,
\begin{align}\label{E1}F^\omega(t):=\int_0^t\int  \frac 1 2 |\omega|^2(u\cdot \nabla \phi )~dx~ds&= \int \frac 1 2 |\omega(x,t)|^2\psi(x)~dx +\int_0^t \int |\nabla \omega|^2\phi ~dx~ds
\notag\\&\qquad-\int_0^t \int \frac 1 2 |\omega|^2(\partial_s \phi +\Delta \phi)~dx~ds -\int_0^t \int (\omega\cdot\nabla)u\cdot (\phi\omega) ~dx~ds
\notag \\&\qquad- \int_0^t \int (b\cdot \nabla)j \cdot (\phi \omega)~dx~ds +\int_0^t\int (j \cdot \nabla)b\cdot (\phi \omega )~dx~ds
\notag \\&=\int \frac 1 2 |\omega(x,t)|^2\psi(x)~dx +\int_0^t \int |\nabla \omega|^2\phi ~dx~ds
\\&\qquad + H^\omega + N_1^\omega+L^\omega+N_2^\omega,\notag
\end{align}while the localized magnetic enstrophy flux is given by,
\begin{align}\label{E2}F^j(t):=\int_0^t\int  \frac 1 2 |j|^2(u\cdot \nabla \phi )~dx~ds&= \int \frac 1 2 |j(x,t)|^2\psi(x)~dx +\int_0^t \int |\nabla j|^2\phi ~dx~ds
\notag\\&\qquad -\int_0^t \int \frac 1 2 |j|^2(\partial_s \phi +\Delta \phi)~dx~ds +\int_0^t \int (\omega\cdot\nabla)b\cdot (\phi j) ~dx~ds
\notag \\&\qquad - \int_0^t \int (b\cdot \nabla)\omega \cdot (\phi j)~dx~ds - \int_0^t\int (j \cdot \nabla)u\cdot (\phi j )~dx~ds
\notag \\&\qquad-\int_0^t\int \bigg(2\sum_{l=1}^3\nabla u_l\times\nabla b_l\bigg)\cdot (\phi j)~dx~ds
\notag \\&=\int \frac 1 2 |j(x,t)|^2\psi(x)~dx +\int_0^t \int |\nabla j|^2\phi ~dx~ds
\\&\qquad + H^j + N_1^j+L^j+N_2^j+X,\notag
\end{align}
and we label their combination as $F(t)=F^\omega(t)+F^j(t)$.

To establish the concentrative effect of inertial forces on the combined enstrophy we will show that the ensemble averages of localized spatio-temporal averages of the above densities associated to an arbitrary $(K_1,K_2)$-cover are positive and nearly constant across a range of scales.
Before formulating our assumptions we specify several technical values.  Fix a value $K_*$ so that, \[K_*\geq \max\{(K_1K_2)^{1/2},3K_2/4,K_1\},\] and set, \[\alpha = 4 K_P K_*^2,\] where $K_P$ is a constant which will be quantified in Section \ref{sec:argument}.

Our assumptions are:
\begin{itemize}
\item[(A1)]{\bf Hybrid Geometric/Smoothness Assumption.}  It is assumed for some threshold $M>0$ that we have,
\begin{align*} |\omega (x+y,t)-\omega (x,t)|\leq |\omega (x+y,t)|{|y|^\frac 1 2},
\end{align*}provided $|y|<2R_0+R_0^\frac 2 3$, $x$ in $\{|\nabla u|>M\}$, and $\omega(x+y)\neq 0$.

\begin{remark}\label{rem:assumptions}{\em This assumption is less satisfying than its analogue in the fluid case of \cite{DaGr3} where it was sufficient to assume the numerically and observationally motivated assumption of {\em coherence of the direction of vorticity} as this depleted the only non-inertial effect (that of vortex stretching).  In our setting, coherence of the vorticity does not deplete all non-inertial effects but (A1) does.  

It is worth mentioning another sufficient formulation as it may be more physically appropriate if the current field is less volatile than the vorticity field.  In the modified formulation we assume the vorticity field satisfies a directional coherence condition identical to that in \cite{DaGr3} and the current satisfies a hybrid geometric/smoothness condition (like (A1) but with $\omega$ replaced with $j$).  The assumption on $j$ depletes all non-linear terms except the vortex stretching term via a kinematic argument which mirrors that given in the next section for $\omega$ while the vortex stretching term is depleted by the coherency assumption. 
This alternative configuration is motivated by the 2D dynamics (cf. \cite{DB}) where the current and vorticity concentrate on largely overlapping regions. Here, the current assumes the structure of an elongated monopole while the vorticity-structure consists of four monopoles squished together in a grid so that the orientation of the vorticity is opposite on adjacent monopoles. Extending this intuition to 3D hints that the current is less oscillatory than the vorticity.}
\end{remark}

\item[(A2)] {\bf Modified Kraichnan-Type Scale.} Let $e_0$, $E_0$, and $P_0$ denote the time averaged total energy, total enstrophy, and modified total palenstrophy at the integral scale. Precisely,
\begin{align*} e_0&= \frac 1 T \int_0^T \frac 1 {R_0^3} \int \phi_0^{4\rho-3}\bigg(\frac {|u|^2} 2 + \frac {|b|^2} 2 \bigg)~dx~ds,
\\E_0&= \frac 1 T \int_0^T \frac 1 {R_0^3} \int \phi_0^{2\rho-1}\big(|\omega|^2+|j|^2)~dx~ds ,
\\P_0&= \frac 1 T\int_0^T \frac 1 {R_0^3} \int \phi_0 \big(|\nabla \omega|^2+|\nabla j|^2)~dx~ds+ \frac 1 {TR_0^3}\int \frac 1 2 |\omega(x,T)|^2 \psi_0(x)~dx.
\end{align*}
The modification of palenstrophy is due to the nature of the temporal cut-off; in addition, note that the cut-off's are modified for technical reasons and $\rho$ was specified in the construction of these functions.

Set,\[\mathcal E_0=\bigg(\frac {E_0} {P_0}\bigg)^{\frac 1 2},\]
and,
\[\varepsilon_0=\bigg( \frac {e_0}{P_0}\bigg)^\frac 1 4.\]
 Define the modified Kraichnan-type scale $\sigma_0$ by,
 \[\sigma_0=\max\{\mathcal E_0, \varepsilon_0\}.\]

Our assumption (A2) is that, for some constant $\beta=\beta(M,K_1,K_2,\int_0^T ||\omega||_2^2~dt)$ with $0<\beta<1$ (the precise value will be identified later), we have, \[\sigma_0<\beta R_0.\]

\begin{remark}{\em The Kraichnan-type scale determines the lower limit of scales at which the concentrative effect is affirmed.  For us this is realized by restricting to scales $R$ with $\sigma_0/\beta<R$.  In comparison to the analogous and identically named parameter in the 3D NSE case we here see a correction of $\beta$ by a power of $1/2$ necessitated by the emergence of energy-level quantities in Section \ref{sec:bounds}.}
\end{remark}

\item[(A3)] {\bf Localization and Modulation} -- Because $\int_0^T ||\omega||_2^2~ds$ is an {\em a priori} bounded quantity (cf. \cite{SeTe}), for a given constant $C_0>0$ there exists $R_0^*$ so that, for any $R_0\leq R_0^*$, we have, \begin{align*} \bigg(\int_0^T||\omega||_{L^2(B(0,2R_0+R_0^\frac 2 3))}^2 ~ds \bigg)^{1/2}\leq \frac 1 {C_0}.
\end{align*} The {\em localization assumption} on $R_0$, the radius of the integral scale, is that, for $C_0=\alpha$, we have $R_0\leq R_0^*$. A precise (up to certain parameters) value for $\alpha$ will materialize in the proof.

The {\em modulation assumption} imposes a restriction on the time evolution of the integral-scale kinetic and magnetic enstrophies across $(0,T)$ consistent with our choice of the temporal cut-off.  Precisely,
\begin{align*}\int |\omega (x,T)|^2\psi_0(x)~dx&\geq \frac 1 2 \sup_s \int |\omega(x,s)|^2\psi_0(x)~dx,
\\ \int |j (x,T)|^2\psi_0(x)~dx&\geq \frac 1 2 \sup_s \int |j(x,s)|^2\psi_0(x)~dx.
\end{align*}

\begin{remark}{\em Regarding localization, we have essentially introduced an upper bound on the range of scales across which the concentrative effect is affirmed and note that this restriction is largely technical (see \cite{DaGr3} for a preliminary statement of a lemma which  will show that the near-constancy across a bounded range of scales extends to a range above that bound).
}\end{remark}

\end{itemize}

Using the terminologies of the above assumptions we are ready to state our main result which establishes the positivity and near-constancy of the combined enstrophy flux across a range of physical scales, the implication of which is that the combined enstrophy is concentrated by inertial forces.

\begin{theorem}\label{Thrm:Cascade} If a weak solution $u,b$ of 3D MHD satisfies (A1)-(A3) on $B(0,2R_0)\times (0,T)$ then, \begin{align*}\frac {1} {4K_*} P_0\leq \langle \Phi  \rangle_R  \leq 4K_* P_0, \end{align*}
for all $\frac {\sigma_0} \beta \leq R\leq R_0$ and $K_*$ dependent only on $K_1$ and $K_2$.
\end{theorem}

\begin{remark}
\emph{It will be plain that the \emph{localization estimates} to be presented in the following section imply that (A1) alone guarantees smoothness over the spatio-temporal integral domain; hence, we are effectively concerned with the global-in-space ($\mathbb{R}^3$) weak solutions that are smooth over the integral domain. However, since we do not impose \emph{any boundary conditions} on the integral domain, the control over the `smooth' norms is \emph{strictly local}.}
\end{remark}

Before continuing to the proof of Theorem \ref{Thrm:Cascade} we observe that enstrophy flux locality -- the transport of the combined enstrophy is  predominantly between scales of comparable size -- is an immediate corollary.  Discussion of this corollary and its precise statement is withheld until Section 6.

To prove Theorem \ref{Thrm:Cascade}, we will confine ensemble averages of the localized (to a ball of radius $R$ centered at $x_0$) combined enstrophy flux between scale- and cover-independent multiples of the localized (to the integral domain) total-palenstrophy, $P_0$.  Local estimates to this effect are contained in Section \ref{sec:bounds}.  Based on these, in Section \ref{sec:argument}, the ensemble averaging methodology is applied to complete the proof of Theorem \ref{Thrm:Cascade}.

\section{Bounds.}\label{sec:bounds}
In this section each of the terms from (\ref{E1}) and (\ref{E2}) are bounded by quantities which can be related to $e_0$, $E_0$, and $P_0$ via the apparatus of ensemble averaging with respect to $(K_1,K_2)$-covers at scale $R$.  Throughout, we limit our consideration to a fixed ball of radius $R$ and suppress the corresponding subscripts.  We label various constants by $K$, $K_e$, $K_E$, and $K_P$ and note these are dependent on $K_1$, $K_2$ and quantities determined by structural properties of 3D MHD.

Bounds for the linear terms in \eqref{E1} and \eqref{E2} follow simply from properties of the spatial cut-off (see \eqref{spacecutoff}):
\begin{align*}H:=H^\omega+ H^j&\leq  \frac {K_E}{R^2 } \int_0^T \int\phi^{2\rho-1}\big(|\omega|^2+|j|^2\big)~dx~ds.
\end{align*}

Before proceeding to bound the nonlinear terms a digression is necessary to introduce the kinematic framework derived in \cite{Co94} and adapted to MHD in \cite{Wu00} and \cite{HeXin2}.  Recall that the deformation tensor of the velocity field can be decomposed in terms of a symmetric component, the strain tensor of $u$, $S$, and a skew component, $\omega\times \cdot $.  Put precisely,
\begin{align}\nabla u =S -\frac 1 2 \omega\times .
\end{align}
The operators in the above decomposition have the following singular integral representations:
\begin{align}\omega(x)&=\frac 1 {4\pi} ~P.V.~\int \sigma(\hat y)\omega(x+y)\frac {dy}{|y|^3},
\end{align}and,
\begin{align}S(x)&=\frac 3 {4\pi} ~P.V.~\int M(\hat y ,\omega(x+y))\frac {dy}{|y|^3},
\end{align}
for, \begin{align*}\hat y =\frac y {|y|}, \sigma(\hat y)=3 \hat y\otimes \hat y -I,~  \mbox{and}~ M(\hat y,f)=\frac 1 2 \big( \hat y \otimes (\hat y \times f)+(\hat y \times f)\otimes \hat y\big).
\end{align*}
A key feature for our treatment of the term $\nabla u_l$ will follow from the fact that $\sigma$ and $M$ (the latter when $f$ is held constant as a function of $y$) have mean zero on the unit sphere.  Integral operators such as these are discussed in \cite{St93} and \cite{St70}.  We connect the above to the term $\nabla u_l\times \nabla b_l$ by noting for the unit vector $e_l$ we have,
\begin{align*}\nabla u_l&=S u~e_l-\frac 1 2 \omega\times e_l.
\end{align*}

Using the zero mean value property we write,
\begin{align*}|\nabla u_l|&\leq \bigg| P.V.\int_{|y|<R^{2/3}} \bigg( \frac 1 {4\pi }\sigma(\hat y) \big(\omega(x+y)-\omega(x)\big)+\frac 3 {4\pi} M(\hat y, \omega(x+y)-\omega (x)) \bigg)\frac {dy} {|y|^3} \bigg|
\\&\qquad + \bigg|\int_{|y|\geq R^{2/3}}\bigg( \frac 1 {4\pi }\sigma(\hat y) \omega(x+y)+\frac 3 {4\pi} M(\hat y, \omega(x+y))\bigg) \frac {dy} {|y|^3}\bigg|
\\&=I_1+I_2.
\end{align*}

Our treatment is now divided between $I_1$ and $I_2$.  For the former, the hybrid geometric-smoothness assumption (A1) entails that,
\begin{align*}I_1&\leq K(\sigma,M)\int_{|y|\leq R^{2/3}}|\omega(x+y)-\omega(x)|\frac {dy } {|y|^3}
\\&\leq  K(\sigma,M)\int_{|y|\leq R^{2/3}}|\omega(x+y)|\frac {dy } {|y|^{5/2}},
\end{align*}and, therefore, by the Hardy-Littlewood-Sobolev inequality (cf. Chapter V of \cite{St93}),
\begin{align}\label{ineq:graduinterior}||I_1||_{3}\leq K ||\omega ||_{L^2(B(x_i,R^{2/3}))}.
\end{align}
Regarding $I_2$, H\"older's inequality allows that,
\begin{align}\notag I_2&\leq K(\sigma, M)\int_{|y|\geq R^{2/3}} \frac 1 {|y|^2} \frac {\omega(x+y)} {|y|}~dy
\\&\leq K(\sigma,M) \bigg( \int_{|y|\geq R^{2/3}} \frac 1 {|y|^4}~dy \bigg)^{\frac 1 2} \bigg(\int_{|y|\geq R^{2/3}} \frac {|\omega(x+y)|^2} {|y|^2}~dy\bigg)^\frac 1 2\notag
\\&\leq {K(\sigma, M)} \frac 1{R^{1/3+2/3}}||\omega||_{L^2(\R^3)}=\frac K R ||\omega||_{L^2(\R^3)}.\label{ineq:graduexterior}
\end{align}

Note that we can apply the exact same argument to $\omega$ alone  to obtain,
\begin{align}\label{ineq:omegaintegralbound}\omega(x)&=\frac 1 {4\pi} P.V. \int_{|y|<R^{2/3}}\sigma(\hat y)(\omega (x+y)-\omega(x))\frac {dy} {|y|^3}+\frac 1 {4\pi}  \int_{|y|\geq R^{2/3}}\sigma(\hat y)\omega (x+y)\frac {dy} {|y|^3}
\\&\leq K ||\omega||_{L^2(B(x_i,R^{2/3})}+\frac K R ||\omega||_{L^2(\R^3)}.\notag
\end{align}

We are now ready to establish bounds on the coupled non-linear terms.  We begin with the most involved, the term involving $\nabla u_l \times \nabla b_l$, that labelled $X$, as this will illustrate many of the computational steps necessary for the other terms. We will show that,
\begin{align}\label{ineq:X}\int_0^T \int \phi j \cdot \nabla u_l\times\nabla b_l ~dx~ds  &\leq \frac {K_P} \alpha \bigg(\frac 1 2 \sup_s ||\psi^\frac 1 2 j||_2^2 + \int_0^T \int \phi |\nabla j|^2~dx~ds\bigg)
\\\notag &\qquad + \frac {K_E} {R^2} \int_0^T \int \phi^{2\rho-1}|j|^2~dx~ds+\frac {\alpha^2K_e} {R^4} \int_0^T \int \phi^{4\rho-3}\frac {|b|^2} {2}~dx~ds.
\end{align}


We begin by splitting the spatial integral into the regions where $|\nabla u|\geq M$ and the complement.  Considering the complement,
\begin{align}\int_0^T \int_{|\nabla u|\leq M} \phi j \cdot \nabla u_l\times\nabla b_l ~dx~ds  &\leq \int_0^T  \int_{|\nabla u|\leq M} M |\phi^{1/2}j||\phi^{1/2}\nabla b_l|~dx~ds
\notag \\&\leq K\frac {M^2} {R^2} \int_0^T \int \phi^{2\rho-1} |j|^2~ds+\frac 1 {R^2} \int_0^T\int \phi^{2\rho-1} |\nabla b_l|^2~ds.
\notag
\end{align} The second integral above can be expressed in terms of energy and palenstrophy level terms.  Indeed, for $k$, $h$, and $l$ distinct elements of $\{1,2,3\}$,
\begin{align}\notag \int_0^T  \int \phi^{2\rho-1} (\partial_i b_l)^2~dx~ds&=-\int_0^T  \int \phi^{2\rho-1} b_l \partial_i\partial_i b_l ~dx~ds -\int_0^T  \int \partial_i \phi^{2\rho-1} b_l \partial_i b_l ~dx~ds
\\\notag&= \int_0^T  \int \phi^{2\rho-1} b_l(\partial_k j_h-\partial_h j_k) ~dx~ds +\frac 1 2 \int_0^T  \int \partial_i\partial_i \phi^{2\rho-1} b_l^2 ~dx~ds
\\\label{Ineq:preAlpha}&\leq 2\int_0^T\int \phi^{2\rho-1}|b||\nabla j|~dx~ds+\frac 1 2 \int_0^T  \int |\partial_i\partial_i \phi^{2\rho-1}| |b|^2 ~dx~ds
\\\label{Ineq:introduceAlpha}&\leq \int_0^T\int \big(2 \alpha^\frac 1 2 \phi^{2\rho-3/2}|b| \big)\big( \alpha^{-\frac 1 2}\phi^{1/2}|\nabla j|\big) ~dx~ds %
\\\notag&\qquad +\frac K {R^2} \int_0^T  \int \phi^{4\rho-3} |b|^2 ~dx~ds.
\end{align}

Applying Young's inequality to the first term of \eqref{Ineq:introduceAlpha} gives us a final bound for the case when $|\nabla u|<M$,
\begin{align}\notag \int_0^T \int_{|\nabla u|\leq M} \phi j \cdot \nabla u_l\times\nabla b_l ~dx~ds  &\leq \frac 1 \alpha \int_0^T \int \phi |\nabla j|^2~dx~ds + \frac {K_E} {R^2} \int_0^T\int\phi^{2\rho-1} |j|^2~dx~ds
\\\notag&\qquad+ \frac {\alpha K_e} {R^4} \int_0^T\int \phi^{4\rho-3}|b|^2~dx~ds.
\\\label{ineq:X1}&\leq \frac 1 \alpha \bigg(\frac 1 2 \sup_s ||\psi^\frac 1 2 j||_2^2 + \int_0^T \int \phi |\nabla j|^2~dx~ds\bigg)
\\\notag&\qquad +\frac {K_E} {R^2} \int_0^T \int \phi^{2\rho-1}|j|^2~dx~ds+\frac {\alpha^2K_e} {R^4} \int_0^T \int \phi^{4\rho-3}\frac {|b|^2} {2}~dx~ds.
\end{align}


Looking now at the regions of high spatial complexity -- i.e. $|\nabla u|\geq M$ -- we split into two cases using the decomposition for $\nabla u_l$,
\begin{align}\notag\bigg|\int_0^T \int_{|\nabla u|\geq M} \phi j \cdot \nabla u_l\times\nabla b_l ~dx~ds  \bigg|&\leq \int_0^T \int_{|\nabla u|\geq M} I_1 |\phi^{1/2}\nabla b_l ||\phi^{1/2}j|~dx~ds
\\\notag &\qquad +\int_0^T\int_{|\nabla u|\geq M} I_2|\phi^{1/2}\nabla b_l ||\phi^{1/2}j|~dx~ds.
\end{align}
For the integral involving $I_1$, the hybrid geometric/smoothness assumption plus the localization and modulation assumptions will serve to minimize palenstrophy level terms.
Applications of H\"older's inequality and the bound \eqref{ineq:graduinterior} and subsequently the Gagliardo-Nirenberg inequality and Young's inequality yields an initial bound,
\begin{align}\notag \int_0^T \int_{|\nabla u|\geq M} I_1 |\phi^{1/2}\nabla b_l ||\phi^{1/2}j|~dx~ds&\leq K \bigg( \int_0^T ||\omega||_{L^2(B(x_i,R^{2/3})}^2~ds\bigg)^{\frac 1 2} \bigg( \int_0^T ||\phi^\frac 1 2 \nabla b_l||_6^2||\phi^{\frac 1 2} j||_2^2~ds\bigg)^\frac 1 2
\\\notag &\leq  \frac 1 {\alpha}  \bigg(\frac 1 2 \sup_{s}||\psi^\frac 1 2 j||_2^2+ K \int_0^T ||\nabla(\phi^\frac 1 2 \nabla b_l)||_2^2~ds\bigg)
\\\label{ineq:superthreshold1}&\leq \frac 1 {\alpha}  \bigg(\frac 1 2 \sup_{s}||\psi^\frac 1 2 j||_2^2+ K \int_0^T ||\nabla \phi^\frac 1 2\otimes \nabla b_l||_2^2~ds
\\&\notag \qquad + K \int_0^T ||\phi^\frac 1 2 \nabla\nabla b_l||_2^2~ds\bigg).
\end{align}where the constant $\alpha$ emerges from the localization assumption, (A3).  We further decompose the last two terms.  For the first, by expanding the integral, using the bound \eqref{Ineq:preAlpha}, applying Young's inequality, and employing the properties of our cut-off functions, we have the following string of bounds:
\begin{align*}\frac K \alpha \int_0^T ||\nabla \phi^\frac 1 2\otimes \nabla b_l||_2^2~ds&=\frac K \alpha \int_0^T \int (\partial_i \phi^\frac 1 2 )^2 (\partial_j b_l)^2~dx~ds
\\&\leq \frac K {\alpha R^2} \int_0^T \int  \phi^{2\rho-1} (\partial_j b_l)^2~dx~ds
\\&\leq \frac K {\alpha R^2}\int_0^T\int \phi^{2\rho-1}|b||\nabla j|~dx~ds+\frac K {\alpha R^2} \int_0^T  \int |\partial_i\partial_i \phi^{2\rho-1}| |b|^2 ~dx~ds
\\&\leq \frac 1 \alpha\int_0^T \int \phi |\nabla j|^2~dx~ds+ \frac K {\alpha R^4}\int_0^T\int \phi^{4\rho-3}|b|^2 ~dx~ds.
\end{align*}

The last term of \eqref{ineq:superthreshold1} can be expressed in a similar fashion as the above but with several additional steps.
To begin, observe that by the product rule and integration by parts, we have the identity,
\begin{align*}||\phi^\frac 1 2 \nabla\nabla b_l||_2^2=\int \phi (\partial_i\partial_j b_l)^2~dx &= \int (\partial_i\partial_j \phi) (\partial_j b_l)(\partial_i b_l)~dx+\int \phi (\partial_j\partial_j b_l)(\partial_i\partial_i b_l)~dx
\\&\qquad+\int (\partial_j\phi )(\partial_j b_l)(\partial_i\partial_i b_l)~dx+ \int (\partial_i \phi) (\partial_j\partial_j b_l)( \partial_i b_l)~dx.
\end{align*}
We next bound each of the terms in the above identity.  For the first, by Young's inequality, we have,
\begin{align*}\frac K \alpha \bigg|\int   (\partial_i\partial_j \phi) (\partial_j b_l)(\partial_i b_l)~dx \bigg|&\leq  K \int |\partial_i\partial_j \phi ||\partial_jb_l||\partial_i b_l|~dx
\\&\leq \frac K {  R^2}\int \phi^{2\rho-1}|\partial_jb_l||\partial_i b_l|~dx
\\&=\frac K {  R^2}\int \bigg(\phi^{\rho-1/2}|\partial_jb_l|\bigg)\bigg(\phi^{\rho-1/2}|\partial_i b_l|\bigg)~dx
\\&\leq \frac K {  R^2} \int \phi^{2\rho-1}|\nabla b_l|^2~dx.
\end{align*}
Moving on, since $b$ is divergence free and assuming $k$, $h$, and $l$ are distinct, we have,
\begin{align*}\frac K \alpha \bigg|\int (\partial_j\phi )(\partial_j b_l)(\partial_i\partial_i b_l)~dx\bigg|&=\frac K \alpha \int |\partial_j \phi||\partial_j b_l|| \partial_h j_k -\partial_k j_h |~dx
\\&\leq \frac K \alpha \int |\partial_j \phi||\partial_j b_l||\partial_h j_k|~dx
\\&\leq \int \bigg( \frac K {\alpha^\frac 1 2 R} \phi^{\rho-1/ 2} |\partial_j b_l|\bigg)\bigg(\frac 1 {\alpha^{\frac {1} 2}} \phi^\frac 1 2 |\partial_h j_k|\bigg)~dx
\\&\leq \frac K {\alpha R^2} \int \phi^{2\rho-1}|\nabla b_l|^2~dx+\frac 1 {2\alpha} \int \phi |\nabla j|^2~dx
\\&\leq  \frac K {R^4} \int \phi^{4\rho-3} |b|^2~dx +\frac 1 \alpha   \int \phi |\nabla j|^2~dx .
\end{align*}
Using Young's inequality and properties of the cut-off function, the remaining term is bounded as,
\begin{align*} \frac K {\alpha }\bigg|\int \phi (\partial_j\partial_j b_l)(\partial_i\partial_i b_l)~dx\bigg|&\leq  \frac K {\alpha }\int \phi  (\partial_i\partial_i b_l)^2 ~dx+ \frac K {\alpha } \int \phi (\partial_j\partial_j b_l)^2 ~dx
\\&\leq \frac {K_P} {\alpha } \int \phi |\nabla j|^2~dx.
\end{align*}

Combining the above (and observing $\alpha>1$) gives a final bound for the term involving $I_1$:
\begin{align}\notag \int_0^T \int_{|\nabla u|\geq M} \phi j \cdot \nabla u_l\times\nabla b_l ~dx~ds  &\leq \frac {K_P }{\alpha} \bigg(\frac 1 2 \sup_{s}||\psi^\frac 1 2 j||_2^2+\int_0^T \int \phi |\nabla j|^2~dx~ds\bigg)
\\\label{ineq:X2} &\qquad +\frac {\alpha K_e} { R^4}\int_0^T\int \phi^{4\rho-3}|b|^2~dx~ds
\end{align}

Turning now to the non-singular case of our decomposition of $|\nabla u_l|$, we again use $\alpha$ to de-emphasize non-localization-apt palenstrophy level quantities.  Its emergence is forced upon an application of Young's inequality with a reciprocal cost to an energy level quantity.  Using the direct estimate \eqref{ineq:graduexterior} on $|I_2|$ and then applying H\"older's inequality, Young's inequality, and the same steps used to obtain and proceed from \eqref{Ineq:introduceAlpha}, we see that,
\begin{align}\notag\int_0^T \int_{|\nabla u|\geq M} I_2 |\phi^{1/2}\nabla b_l ||\phi^{1/2}j|~dx~ds&\leq \frac K {R} \bigg(\int_0^T ||\omega||_2^2~ds\bigg)^{\frac 1 2} \bigg(  \int_0^T ||\phi^\frac 1 2j||_2^2||\phi^\frac 1 2 \nabla b_l||_2^2~ds\bigg)^\frac 1 2
\\\notag&\leq \bigg(\frac 1 {\sqrt \alpha}\sup_s ||\psi^\frac 1 2 j||_2\bigg)\bigg( \frac { \alpha K} {R^2} \int_0^T ||\phi^\frac 1 2 \nabla b_l||_2^2 ~ds\bigg)^\frac 1 2
\\\notag&\leq \frac 1 \alpha \frac 1 2 \sup_s ||\psi^\frac 1 2 j||_2^2 + \frac {\alpha K} {R^2}  \int_0^T ||\phi^\frac 1 2 \nabla b_l||_2^2 ~ds
\\\label{ineq:X3}&\leq  \frac 1 \alpha \bigg(\frac 1 2 \sup_s ||\psi^\frac 1 2 j||_2^2 + \int_0^T \int \phi |\nabla j|^2~dx~ds\bigg)
\\\notag&\qquad+ \frac {\alpha K_e} {R^4} \int_0^T \int \phi^{4\rho-3} \frac {|b|^2} 2 ~dx~ds.
\end{align}

Combining the bounds \eqref{ineq:X1}, \eqref{ineq:X2}, and \eqref{ineq:X3} establishes the bound \eqref{ineq:X} and concludes our discussion of the term involving $\nabla u_l\times \nabla b_l$.


Bounds for the remaining four critical order nonlinear terms, $N_1^\omega$, $N_2^\omega$, $N_1^j$, and $N_2^j$, are now attended. The processes for estimating $N_1^\omega$ and $N_2^j$ are identical up to labeling. We only illustrate the latter. Splitting the space integral, when $|\nabla u|<M$, we have,
\begin{align*}N_2^j=\int_0^T\int_{|\nabla u|<M} (j\cdot \nabla)u\cdot \phi j ~dx~ds&\leq M\int_0^T ||\phi^\frac 1 2 j||_2^2 ~ds.
\end{align*}
When $|\nabla u|\geq M$, noting by direct comparison that $|j|\leq \sqrt 5 |\nabla b|$, we bound the quantity, \[\int_0^T\int_{|\nabla u|\geq M}|\nabla u||\phi^\frac 1 2 \sqrt 5\nabla b||\phi^\frac 1 2j|~dx~ds,\]
in an identical fashion to the unique-to-MHD term.  The resultant final bounds are,
\begin{align}N_2^j=\int_0^T\int (j\cdot \nabla)u\cdot \phi j~dx~ds&\leq \frac {K_P} \alpha \bigg(\frac 1 2 \sup_s ||\psi^\frac 1 2 j||_2^2 + \int_0^T \int \phi |\nabla j|^2~dx~ds\bigg)
\notag\\\label{ineq:N2b}&\qquad + \frac {K_E} {R^2} \int_0^T \int \phi^{2\rho-1}|j|^2~dx~ds+\frac {\alpha^2 K_e} {R^4} \int_0^T \int \phi^{4\rho-3}\frac {|b|^2} {2}~dx~ds,
\end{align}and,
\begin{align}\notag N_1^\omega=\int_0^T\int (\omega\cdot\nabla)u\cdot \phi\omega~dx~ds&\leq \frac {K_P} \alpha \bigg(\frac 1 2 \sup_s ||\psi^\frac 1 2 \omega||_2^2 + \int_0^T \int \phi |\nabla \omega|^2~dx~ds\bigg)
\notag\\\label{ineq:N1u}&\qquad + \frac {K_E} {R^2} \int_0^T \int \phi^{2\rho-1}|\omega|^2~dx~ds+\frac {\alpha^2 K_e} {R^4} \int_0^T \int \phi^{4\rho-3}\frac {|u|^2} {2}~dx~ds.
\end{align}

The space-time integrals of $N_1^j$ and $N_2^\omega$ also enjoy formally identical bounding procedures. The spatial integrals are still split into regions depending on $|\nabla u|$.  When $|\nabla u|<M$ we have the point-wise estimate $|\omega|\leq 5^{1/2} M$ and we obtain a scaled version of prior bounds.  When $|\nabla u|\geq M$ we use the kinematic estimate, \eqref{ineq:omegaintegralbound}, on $|\omega|$.  The same familiar argument now shows that $N_1^j$ and $N_2^\omega$ are bounded by the same dominating quantity appearing in \eqref{ineq:N2b}.

To summarize,
\begin{align}\notag N:=N_1^\omega+N_2^\omega+N_1^j+N_2^j&\leq \frac {K_P} \alpha \bigg(\frac 1 2 \sup_s \big(||\psi^\frac 1 2 \omega||_2^2+||\psi^\frac 1 2 j||_2^2\big) + \int_0^T \int \phi \big( |\nabla \omega|^2+|\nabla j|^2\big)~dx~ds\bigg)
\\\notag&\qquad+\frac {K_E} {R^2} \int_0^T \int \phi^{2\rho-1}\big(|\omega|^2+|j|^2\big)~dx~ds
\\\notag &\qquad+\frac {\alpha^2 K_e} {R^4} \int_0^T \int \phi^{4\rho-3} \frac {|u|^2+|b|^2} {2}~dx~ds.\end{align}

Examining now $L^\omega$ and $L^j$, we note, upon integrating by parts, a cancellation occurs when we consider their sum: \[L:=L^\omega+L^j =  \int_0^T\int (j\cdot\omega)(\nabla \phi \cdot b)  ~dx~ds. \]Although the above term is lower order and can be bounded in a more efficient way than what transpires below, we choose the less direct approach to limit problem specific dependencies of the parameter $\beta$.  We again split the spatial integral to obtain on one hand that,
\begin{align*}\int_0^T\int_{|\nabla u|\leq M} (j\cdot\omega)(\nabla \phi \cdot b)  ~dx~ds&\leq \sqrt 5 M \frac 1 R \int_0^T ||\phi^\frac 1 2 b||_2||\phi^{\rho-1/2} j||_2 ~ds
\\&\leq \frac {K_E} {R^2}\int_0^T\int \phi^{2\rho-1}|j|^2~dx~ds + \frac {\alpha^2 K_e} {R^4} \int_0^T\int \phi^{4\rho-3}|b|^2~dx~ds.\end{align*}
On the other hand, using the kinematic decomposition of $\omega$ as a singular integral, we have,
\begin{align*}\int_0^T\int_{|\nabla u|\geq M} \int_{|y|\geq R^{2/3}}\sigma (\hat y)\omega(x+y)~\frac {dy}{|y|^3} |j||\nabla \phi \cdot b|  ~dx~ds&\leq \frac 1 \alpha \frac 1 2 \sup_s ||\psi^\frac 1 2 j||_2^2
\\&\qquad+ \frac {\alpha^2 K_e} {R^4} \int_0^T\int \phi^{4\rho-3} |b|^2~dx~ds,
\end{align*}
and,
\begin{align}\notag\int_0^T\int_{|\nabla u|\geq M} \int_{|y|< R^{2/3}}\sigma (\hat y)\omega(x+y)~\frac {dy}{|y|^3} |j||\nabla \phi \cdot b|  ~dx~ds&\leq \frac 1 {\alpha}\bigg(\int_0^T \big(||\phi^\frac 1 2 j||_2||\,|\nabla \phi|^\frac 1 2 b||_2\big)^2~dx~ds\bigg)^\frac 1 2
\\\notag&\leq \frac 1 {\alpha} \bigg(\frac 1 2 \sup_s ||\psi^\frac 1 2 j||_2^2 + \int_0^T ||\phi^\frac 1 2 \nabla j||_2^2~dx~ds\bigg)
\\\notag&\qquad + \frac {\alpha^2 K_e} {R^4}\int_0^T \int \phi^{4\rho-3} |b|^2~dx~ds.
\end{align}
Combining the above we conclude that $L$ is also bounded by the quantity given in \eqref{ineq:N2b}.


\section{Proof of the Main Result.}\label{sec:argument}
The proof of Theorem \ref{Thrm:Cascade} is now given.  We work in the context of an arbitrary $(K_1,K_2)$-cover at scale $R$, $\{B(x_i,R)\}_{i=1}^n$, of the integral domain $B(0,R_0)$ where $R<R_0<1$ and assume the premises of Theorem \ref{Thrm:Cascade} hold.

First we establish bounds for averages associated to an arbitrary cover element centered at $x_i$ (note that $x_0$ was arbitrary in the previous subsection and the bounds were independent of $x_0$; our subscripts here indicate localization around $x_i$ at scale $R$):
\begin{align*}\int_0^T \Phi_{x_i,R} ~ds&=\int \frac 1 2 |\omega(x,T)|^2\psi_i(x)~dx +\int_0^T \int |\nabla \omega|^2\phi_i ~dx~ds
\\&\quad+\int \frac 1 2 |j(x,T)|^2\psi_i(x)~dx +\int_0^T \int |\nabla j|^2\phi_i ~dx~ds
\\&\quad+H_i+N_i+L_i+X_i.
\end{align*}
Observe that,
\begin{align*}H_i+N_i+L_i+X_i & \leq \frac {K_P } {\alpha} \bigg( \frac 1 2 \bigg(\sup_s ||\psi_i^\frac 1 2 j||_{L^2(B(x_i,2R))}+\sup_s||\psi_i^\frac 1 2\omega||_{L^2(B(x_i,2R))}\bigg)
\\&\qquad + \int_0^T \int \phi_i\big( |\nabla j|^2+|\nabla \omega|^2\big)~dx~ds\bigg)
\\&\qquad + \frac {K_E} {R^2}  \int_0^T \int \phi_i^{2\rho-1}\big( |\omega|^2+ |j|^2\big) ~dx ~ds
\\&\qquad + \frac {\alpha^2 K_e} {R^4}\int_0^T  \int \phi_i^{4\rho-3} \bigg( \frac{ |b|^2} 2 + \frac{ |u|^2} 2 \bigg)~dx~ds.
\end{align*}

The properties of $(K_1,K_2)$-covers (see \eqref{ineq:interp}) allow us, upon taking ensemble averages and applying the modulation part of (A3), to pass to a lower bound involving only integral scale quantities,
\begin{align*}\bigg\langle \frac 1 T\int_0^T \frac 1 {R^3} \Phi_{x_i,R}~ds \bigg\rangle_R&\geq \frac 1 {K_1} P_0 - K_2\frac {K_P} {\alpha}  P_0-\frac {K_2 K_E} {R^2} E_0 - \frac {\alpha^2 K_2 K_e} {R^4}e_0.
\end{align*}
At this point we specify the value for $\alpha$, \[\alpha= 4K_P K_*^2,\] and recall that, \[K_*\geq\max\{(K_1K_2)^{1/2},3K_2/4,K_1\}.\] Consequently,
\[K_2\frac {K_P} {\alpha}\leq \frac {1}{4K_1},\]and, therefore,
\begin{align*}\bigg\langle \frac 1 T\int_0^T \frac 1 {R^3} \Phi_{x_i,R} ~ds \bigg\rangle_R&\geq \frac 3 {4K_1} P_0 -
\frac {K_2 K_E} {R^2} E_0 - \frac {\alpha^2 K_2 K_e} {R^4}e_0.
\end{align*}

Recalling now (A2) we have,
\begin{align*}\frac {\alpha^2 K_2 K_e} {R^4}e_0\leq \beta^4 \alpha^2 K_2 K_e P_0,
\end{align*}and,
\begin{align*}\frac {K_2  K_E} {R^2}E_0\leq \beta^2 K_2  K_E P_0.
\end{align*}
Therefore, noting $0<\beta<1$,
\begin{align*}\bigg\langle \frac 1 T\int_0^T \frac 1 {R^3} \Phi_{x_i,R} ~ds \bigg\rangle_R&\geq \frac 3 {4K_1} P_0 - {\beta^2 K_2(K_E+\alpha^2 K_e)} P_0.
\end{align*}
The parameter modifying our Kraichnan-type scale, $\beta$, is now chosen so that $\beta^2$ is small enough to satisfy, \[2\beta^2 K_1K_2(K_E+\alpha^2 K_e))<1.\]  Granted this, we obtain,
\begin{align*}\bigg\langle \frac 1 T\int_0^T \frac 1 {R^3} \Phi_{x_i,R} ~ds \bigg\rangle_R&\geq \frac 1 {4K_1} P_0 \geq \frac 1 {4K_*} P_0.
\end{align*}

Establishing the upper bound is, by properties of $(K_1,K_2)$-covers, immediate because we chose $K_*$ so that $4K_*\geq 3K_2$. Indeed,
\begin{align*}\bigg\langle \frac 1 T\int_0^T \frac 1 {R^3} \Phi_{x_i,R} ~ds \bigg\rangle_R&\leq K_2P_0+2P_0\leq 4K_* P_0.
\end{align*}
Having established that,
\begin{align*}\frac 1 {4K_*} P_0\leq \langle \Phi  \rangle_R \leq 4K_* P_0,
\end{align*}the proof is complete.

\section{Flux Locality.} As mentioned previously, we can immediately deduce locality of the flux from Theorem \ref{Thrm:Cascade}.  Flux locality, in the context of turbulence phenomenology, refers to the fact that the transfer takes place primarily between comparable scales.  This is realized in terms of flux by the proposition that the time averaged flux at scale $R$ is well-correlated only with the time averaged fluxes at comparable scales.  While flux locality is phenomenologically accepted in the hydrodynamic case, there has been some controversy about the locality in plasma turbulence (cf. \cite{AlDiss, AlEy10} for a locality result on the energy level, as well as a discussion and references on the topic). The present result establishes that the combined enstrophy flux is local. 

Define the time-averaged local magnetic and kinetic enstrophy fluxes associated to the cover element around the point $x_i$ as follows,
\begin{align*}\Psi_{x_i,R}^\omega&=\frac 1 T \int_0^T \int \frac 1 2 |\omega|^2 (u\cdot \nabla \phi_i)~dx~ds=R^3\Phi_{x_i,R}^\omega,
\\ \Psi_{x_i,R}^j&=\frac 1 T \int_0^T \int \frac 1 2 |j|^2 (u\cdot \nabla \phi_i)~dx~ds=R^3\Phi_{x_i,R}^j,
\end{align*}and, correspondingly, the combined enstrophy flux as $\Psi_{x_i,R}=\Psi_{x_i,R}^\omega+\Psi_{x_i,R}^j$.

 Further, define the ensemble average over a $(K_1,K_2)$-cover of the time-averaged combined flux as, \begin{align*}\langle \Psi  \rangle_R&=\frac 1 n \sum_{i=1}^n \Psi_{x_i,R} =R^3 \langle \Phi  \rangle_R.
\end{align*}
Using the clear relationships between the spatio-temporal and ensemble averaged terms and the time and ensemble averaged terms one can use the bounds established in Theorem \ref{Thrm:Cascade} to directly verify the following theorem (for which the proof is omitted).
\begin{theorem}Let $u$ and $b$ satisfy the assumptions of Theorem \ref{Thrm:Cascade} and let $R$ and $r$ be two scales in the range  $\sigma_0/\beta\leq r,R \leq R_0$.  Then,
\begin{align*}\frac 1 {16K_*^2} \bigg(\frac r R \bigg)^3 \leq \frac {\langle \Psi  \rangle_r} {\langle \Psi  \rangle_R} \leq 16K_*^2\bigg(\frac r R \bigg)^3.
\end{align*}\end{theorem}

\begin{remark}
\emph{Note that along the \emph{dyadic scale} -- $r=2^k R$ -- the locality propagates \emph{exponentially}.}
\end{remark}

\subsection*{Acknowledgements.}

Z.B. acknowledges support of the \emph{Virginia Space Grant Consortium} via the Graduate Research Fellowship; Z.G. acknowledges support of the \emph{Research Council of Norway} via the grant 213474/F20 and the \emph{National Science Foundation} via the grant DMS 1212023.

\bibliographystyle{plain}
\bibliography{refs}

\begin{thebibliography}{10}

\bibitem{AlDiss}
H.~Aluie.
\newblock {\em Hydrodynamic and Magnetohydrodynamic Turbulence: Invariants,
  cascades, and locality.}
\newblock Phd thesis, Johns Hopkins University, 2009.

\bibitem{AlEy10}
H.~Aluie and G.~L. Eyink.
\newblock Scale locality of magnetohydrodynamic turbulence.
\newblock {\em Phys. Rev. Lett.}, 104(8):081101, 2010.

\bibitem{BeBe}
H.~Beir{\~a}o~da Veiga and L.~C. Berselli.
\newblock On the regularizing effect of the vorticity direction in
  incompressible viscous flows.
\newblock {\em Differential Integral Equations}, 15(3):345, 2002.

\bibitem{Be12}
A.~Beresnyak.
\newblock Basic properties of magnetohydrodynamic turbulence in the inertial
  range.
\newblock {\em Mon. Not. R. Astron. Soc.}, 422(4):3495, 2012.

\bibitem{BeLa08}
A.~Beresnyak and Lazarian A.
\newblock Strong imbalanced turbulence.
\newblock {\em ApJ}, 682(2):1070, 2010.

\bibitem{DB}
D.~Biskamp.
\newblock {\em Magnetohydrodynamic Turbulence}.
\newblock Cambridge University Press, 2003.

\bibitem{DB-ES-01}
D.~Biskamp and E.~Schwartz.
\newblock On two-dimensional magnetohydrodynamic turbulence.
\newblock {\em Phys. Plasmas}, 8(7):3282, 2001.

\bibitem{B05}
S.~Boldyrev.
\newblock On the spectrum of magnetohydrodynamic turbulence.
\newblock {\em Astrophys. J. Lett.}, 626(1):L37, 2005.

\bibitem{Borovsky08}
J.~E. Borovsky.
\newblock Flux tube texture of the solar wind: Strands of the magnetic carpet
  at 1 au?
\newblock {\em J. of Geophys. Res.}, 113(A):8, 2008.

\bibitem{BrGr1}
Z.~Bradshaw and Z.~Gruji\'c.
\newblock Energy-level turbulent cascades in physical scales of 3d
  incompressible plasma.
\newblock {\em submitted}, arxiv:1211.3083.

\bibitem{Bruno2001}
R.~Bruno, V.~Carbone, P.~Veltri, E.~Pietropaolo, and B.~Bavassano.
\newblock Identifying intermittency events in the solar wind.
\newblock {\em Planetary and Space Science}, 49(12):1201, 2001.

\bibitem{Chorin}
A.~J. Chorin.
\newblock {\em Vorticity and Turbulence}.
\newblock Number v. 103 in Applied Mathematical Sciences. Springer, 1994.

\bibitem{Co94}
P.~Constantin.
\newblock Geometric statistics in turbulence.
\newblock {\em SIAM Rev.}, 36(1):73, 1994.

\bibitem{CoFe93}
P.~Constantin and C.~Fefferman.
\newblock Direction of vorticity and the problem of global regularity for the
  {N}avier-{S}tokes equations.
\newblock {\em Indiana Univ. Math. J.}, 42(3):775, 1993.

\bibitem{CoPrSe-95}
P.~Constantin, I.~Procaccia, and D.~Segel.
\newblock On the creation and dynamics of vortex structures in
  three-dimensional turbulence.
\newblock {\em Phys. Rev. E}, 51(4):3207, 1995.

\bibitem{DaGr1}
R.~Dascaliuc and Z.~Gruji{\'c}.
\newblock Energy cascades and flux locality in physical scales of the 3{D}
  {N}avier-{S}tokes equations.
\newblock {\em Comm. Math. Phys.}, 305(1):199, 2011.

\bibitem{DaGr3}
R.~Dascaliuc and Z.~Gruji{\'c}.
\newblock Coherent {V}ortex {S}tructures and 3{D} {E}nstrophy {C}ascade.
\newblock {\em Comm. Math. Phys.}, 317(2):547, 2013.

\bibitem{Eyink2006}
G.~L. Eyink and H.~Aluie.
\newblock The breakdown of {A}lfv\'en's theorem in ideal plasma flows:
  Necessary conditions and physical conjectures.
\newblock {\em Phys. D}, 223(1):82, 2006.

\bibitem{Frisch05}
U.~Frisch.
\newblock {\em Turbulence}.
\newblock Cambridge University Press, Cambridge, 1995.
\newblock The legacy of A. N. Kolmogorov.

\bibitem{GaPoMa05}
S.~Galtier, .~Pouquet, and A.~Mangeney.
\newblock On spectral scaling laws for incompressible anisotropic
  magnetohydrodynamic turbulence.
\newblock {\em Phys. Plasmas}, 12(9):092310, 2005.

\bibitem{GS94}
P.~Goldreich and S.~Sridhar.
\newblock Toward a theory of interstellar turbulence. i. weak {A}lfv\'enic
  turbulence.
\newblock {\em Astrophys. J.}, 432:612, 1994.

\bibitem{GS95}
P.~Goldreich and S.~Sridhar.
\newblock Toward a theory of interstellar turbulence. ii. strong {A}lfv\'enic
  turbulence.
\newblock {\em Astrophys. J.}, 438:763, 1995.

\bibitem{Greco08}
A.~Greco, P.~Chuychai, W.~H. Matthaeus, W.~Servidio, and P.~Dmitruk.
\newblock Intermittent mhd structures and classical discontinuities.
\newblock {\em Geophys. Res. Lett.}, 35(L):19111, 2008.

\bibitem{Greco09}
A.~Greco, W.~H. Matthaeus, W.~Servidio, P.~Chuychai, and P.~Dmitruk.
\newblock Statistical analysis of discontinuities in solar wind ace data and
  comparison with intermittent mhd turbulence.
\newblock {\em Astrophys. J.}, 691(L):111, 2009.

\bibitem{HeXin2}
C.~He and Z.~Xin.
\newblock On the regularity of weak solutions to the magnetohydrodynamic
  equations.
\newblock {\em J. Differential Equations}, 213(2):235, 2005.

\bibitem{Ir}
P.~Iroshnikov.
\newblock Turbulence of a conducting fluid in a strong magnetic field.
\newblock {\em Sov. Astron.}, 7:566, 1964.

\bibitem{kinney95}
R.~Kinney, J.~C. McWilliams, and T.~Tajima.
\newblock Coherent structures and turbulent cascades in two-dimensional
  incompressible magnetohydrodynamic turbulence.
\newblock {\em Phys. Plasmas}, 2(10):3623, 1995.

\bibitem{Kr}
R.~H. Kraichnan.
\newblock Inertial range spectrum in hydromagnetic turbulence.
\newblock {\em Phys. Fluids}, 8(7):1385, 1966.

\bibitem{MaMoWaSe-Review}
W.~H. Matthaeus, D.~C. Montgomery, M.~Wan, and S.~Servidio.
\newblock A review of relaxation and structure in some turbulent plasmas:
  magnetohydrodynamics and related models.
\newblock {\em Journal of Turbulence}, 13(37):1, 2012.

\bibitem{MuGr05}
W.C. M\"uller and R.~Grappin.
\newblock Spectral energy dynamics in magnetohydrodynamic turbulence.
\newblock {\em Phys. Rev. Lett.}, 95:114502, Sep 2005.

\bibitem{PeBo10}
J.~Perez and S.~Bolydrev.
\newblock Strong magnetohydrodynamic turbulence with cross helicity.
\newblock {\em Phys. Plasmas}, 17(055903):1, 2010.

\bibitem{PeMaBoCa12}
J.~C. Perez, J.~Mason, S.~Boldyrev, and F.~Cattaneo.
\newblock On the energy spectrum of strong magnetohydrodynamic turbulence.
\newblock {\em Phys. Rev. X}, 2:041005, Oct 2012.

\bibitem{BhaPo10}
J.~J. Podesta and A.~Bhattacharjee.
\newblock Theory of incompressible magnetohydrodynamic turbulence with
  scale-dependent alignment and cross-helicity.
\newblock {\em Astrophys. J.}, 718(2):1151, 2010.

\bibitem{PoBhGa}
D.~Pontin, A.~Bhattacharjee, and K.~Galsgaard.
\newblock Current sheet formation and nonideal behavior at three-dimensional
  magnetic null points.
\newblock {\em Phys. Plasmas}, 14(052106):1, 2007.

\bibitem{SeTe}
M.~Sermange and R.~Temam.
\newblock Some mathematical questions related to the {MHD} equations.
\newblock {\em Comm. Pure Appl. Math.}, 36(5):635, 1983.

\bibitem{Servidio_et_al_2011}
S.~Servidio, P.~Dmitruk, A.~Greco, M.~Wan, S.~Donato, P.~A. Cassak, M.~A. Shay,
  V.~Carbone, and W.~H. Matthaeus.
\newblock Magnetic reconnection as an element of turbulence.
\newblock {\em Nonlinear Processes in Geophysics}, 18(5):675--695, 2011.

\bibitem{St70}
E.~M. Stein.
\newblock {\em Singular Integrals}.
\newblock Princeton Univ. Press, Princeton, 1970.

\bibitem{St93}
E.~M. Stein.
\newblock {\em Harmonic analysis: real-variable methods, orthogonality, and
  oscillatory integrals}, volume~43 of {\em Princeton Mathematical Series}.
\newblock Princeton University Press, Princeton, NJ, 1993.

\bibitem{Wu00}
J.~Wu.
\newblock Analytic results related to magneto-hydrodynamic turbulence.
\newblock {\em Phys. D}, 136(3-4):353, 2000.

\bibitem{Wu2002}
J.~Wu.
\newblock Bounds and new approaches for the 3{D} {MHD} equations.
\newblock {\em J. Nonlinear Sci.}, 12(4):395, 2002.

\bibitem{Zhdankin_et_al_2012}
V.~Zhdankin, S.~Boldyrev, and J.~Mason.
\newblock Distribution of magnetic discontinuities in the solar wind and in
  magnetohydrodynamic turbulence.
\newblock {\em The Astrophysical Journal Letters}, 760(2):L22, 2012.

\end{thebibliography}

\end{document}